 \documentclass[twocolumn]{aastex62}

\acceptjournal{PASP}

\shorttitle{Survey Design for Planets Transiting L and T Dwarfs}
\shortauthors{Tamburo \& Muirhead}

\usepackage{wrapfig}
\usepackage{footnote}

\begin{document}

\title{Design Considerations for a Ground-Based Search for Transiting Planets around L and T Dwarfs}
\correspondingauthor{Patrick Tamburo}
\email{tamburop@bu.edu}

\author[0000-0003-2171-5083]{Patrick Tamburo}

\author[0000-0002-0638-8822]{Philip S. Muirhead}
\affiliation{Department of Astronomy \& The Institute for Astrophysical Research, Boston University, 725 Commonwealth Ave., Boston, MA 02215, USA}

\begin{abstract}


We present design considerations for a ground-based survey for transiting exoplanets around L and T dwarfs, spectral classes that have yet to be thoroughly probed for planets. We simulate photometry for L and T targets with a variety of red-optical and near-infrared detectors, and compare the scatter in the photometry to anticipated transit depths. Based on these results, we recommend the use of a low-dark-current detector with \textit{H}-band NIR photometric capabilities. We then investigate the potential for performing a survey for Earth-sized planets for a variety of telescope sizes. We simulate planetary systems around a set of spectroscopically confirmed L and T dwarfs using measured M dwarf planet occurrence rates from \textit{Kepler}, and simulate their observation in surveys ranging in duration from 120 to 600 nights, randomly discarding 30\% of nights to simulate weather losses. We find that an efficient survey design uses a 2-meter-class telescope with a NIR instrument and 360-480 observing nights, observing multiple L and T targets each night with a dithering strategy. Surveys conducted in such a manner have over an 80\% chance of detecting at least one planet, and detect around 2 planets, on average. The number of expected detections depends on the true planet occurrence rate, however, which may in fact be higher for L and T dwarfs than for M dwarfs.

\end{abstract}



\section{Introduction} \label{sec:intro}

Transit surveys have delivered a wealth of information about the demographics of extrasolar worlds in the past two decades. In the search for systems that resemble our own solar system, many transit surveys have been designed with the goal of determining the planet population around bright, Sun-like main sequence stars. For example, the primary science goal of the \textit{Kepler} mission was to measure the occurrence rate of Earth-sized planets in and around the habitable zones of Sun-like stars  \citep{Borucki2010}. SuperWASP , the most successful ground-based transit survey by number of detected planets, was designed to obtain 1$\%$ photometry on targets brighter than V $\sim$11.5 \citep{Pollacco2006}. The ongoing \textit{TESS} mission will continue this trend, as it searches for transits around 200,000 of the brightest stars in the solar neighborhood at visible wavelengths \citep{Ricker2015}. 

M dwarfs have also received significant attention in transit surveys, largely because their small radii and drawn-in habitable zones make them more suitable for finding and characterizing habitable, Earth-like exoplanets with current and upcoming technologies \citep[e.g.,][]{Morley2017}. In addition, the larger transit depths of Earth-sized planets around M dwarfs enables their detection from the ground. The ground-based MEarth survey \citep{Nutzman2008} targets M dwarfs specifically, and has discovered four planets to date, including the super-Earth GJ 1214 b \citep{Charbonneau2009}, the Earth-sized planet GJ 1132 b \citep{BertaThompson2015}, and two terrestrial planets around LHS 1140 \citep{Dittmann2017,Ment2019}. The TRAPPIST-UCDTS \citep{Gillon2013} program, another ground-based survey, monitored 50 ultra-cool stars (M6- to M9-type), and detected the profoundly impactful TRAPPIST-1 system with seven transiting Earth-sized planets \citep{Gillon2016,Gillon2017}. This program served as a prototype survey for SPECULOOS, which aims to monitor around 1200 targets later than M7 over 10 years to search for habitable, Earth-sized planets  \citep{Burdanov2017,Delrez2018}.

Beyond the M spectral type are the L and T dwarfs (hereafter LTs), which consist of main-sequence stars, brown dwarfs, and planetary-mass objects \citep{Kirkpatrick1999, Burgasser2006}. This ambiguity can be resolved by estimation of an object's age through comparison with evolutionary models \citep[e.g.,][]{Burrows1997}, or using ages of any stellar companions or clusters in which the object resides. Alternatively, the nature of the object can be determined by obtaining direct mass measurements in binary systems \citep[e.g.,][]{Dupuy2014}. Because they can be planetary-mass, LTs represent the very lower bound on what we might consider conventional planetary systems; planets around the lowest mass LTs (those below the deuterium-burning limit) might even be better classified as moons. The study of LTs in the context of their planet populations is therefore a crucial component to developing a general theory of \textit{satellite} formation, and a dedicated search for planets around them will help answer the following questions: Do planet occurrence rates around LTs differ significantly from those around larger stars? Is there any difference in the occurrence rate of planets between stars, brown dwarfs, and planetary-mass objects? Are their planets predominantly terrestrial? Do their terrestrial planets retain detectable atmospheres? 

Frequently referred to with the blanket term ``ultracool dwarfs'' \citep[i.e., objects of type M7 and later,][]{Kirkpatrick1997}, LTs possess effective temperatures ranging from about 500 to 2200 K \citep{Kirkpatrick1999, Burrows2011}. Being supported by electron degeneracy pressure, all LTs are around 1 R$_{Jup}$ in size. This is about half the size of a typical MEarth target \citep[e.g.,][]{Irwin2011}, leading to transit depths that are about four times deeper. Whereas MEarth was designed to detect 2R$_\oplus$ planets \citep{Nutzman2008}, the small radii of LTs facilitate the discovery of transiting Earth-analogs (though M dwarfs are  intrinsically brighter, increasing photometric signal-to-noise). An Earth-radius planet creates a transit depth of about 1\% around an average LT, comparable to that of a hot Jupiter, the first class of transiting exoplanet discovered and successfully characterized \citep[e.g., HD 209458 b,][]{Charbonneau2000,Charbonneau2002}. 

The habitable zones (HZs) of LTs can be efficiently probed for transiting planets due to their low effective temperatures. A typical L0, with a radius of 1 R$_{Jup}$ and T$_{eff}$ = 2200 K, produces a luminosity of $\sim$1/5000 L$_\odot$. A planet orbiting such a host receives the same flux as Earth if it orbits every 1.65 days, which equates to an a/R$_*$ of about 6 to 27 (assuming zero eccentricity, and a host mass range of 1-100 M$_{Jup}$). Just a few nights of observation are needed to search for habitable planets around such a host. Tidal heating, which impacts the energy budget of close-in solar system satellites like Europa \citep[e.g.,][]{Ross1987}, Enceladus \citep[e.g.,][]{Ross1989}, and Io \citep[e.g.,][]{Yoder1979}, may increase the surface temperatures of short-period LT exoplanets, resulting in a HZ further from the host.  

Because the transit method is geometrically biased to finding close-in planets, and because LTs possess drawn-in HZs, planets found through the transit method around LTs are more likely to be in the HZ than those found transiting hotter spectral types. However, it should be noted that the HZ around substellar objects changes significantly over the course of the object's lifetime. Brown dwarfs, unable to sustain hydrogen fusion in their cores, cool significantly over Gyr timescales, causing their HZs to shift inward. For this reason, planets found in the HZ around substellar LTs were probably previously \textit{interior} to the HZ, which could have evaporated any water on the planet through H$_2$O photolysis and thermal escape of hydrogen \citep{Barnes2013}. \citet{Bolmont2017} investigated energy-limited water escape on planets orbiting cooling brown dwarfs, and found that there exist large regions of semi-major axis and host mass parameter space where planets lose less than 1 EO$_H$ (Earth ocean equivalent content of hydrogen) and spend more than 1 Gyr in the HZ, implying that habitability may be maintained in ``sweet spots'' around LTs. 

\subsection{Evidence for Planets around LTs}
An abundance of observational and theoretical evidence suggests that LTs host planets. For one, circumstellar disks have been observed around late M-type brown dwarfs. By identifying disk candidates through excess infrared emission, \citet{Luhman2012} found evidence of an increase in the disk-harboring fraction of members of Upper Scorpius with spectral types later than M5, suggesting that later spectral types retain their disks longer than earlier spectral types. This allows more time for planet formation compared to earlier spectral types. Similar results were found for 15 brown dwarfs in the 25 Orionis group by \citet{Downes2015}, with evidence for disk structure in 33$^{+10.8}_{-9.8}\%$ of sub-stellar objects, versus 3.9$^{+2.4}_{-1.6}\%$ of low-mass stars. 

The simple presence of disks, however, does not automatically imply planet formation; if they contain too little mass, Earth-sized planets may not form. \citet{Payne2007} investigated the ability for brown dwarfs to form planets via core-accretion by extending a semi-analytic model first described in \citet{Ida2004} to the sub-stellar mass regime. Their results suggest that LTs should host planets that are both terrestrial and close-in. Their simulations were unable to produce giant planets, and they found that the radial distribution of planets was a factor of 10 times closer to the host than around Sun-like stars. They concluded that brown dwarfs with disks of a few M$_{Jup}$ should be capable of forming planets up to 5 M$_{\oplus}$ in mass, with separations of a tenth to a few AU; however, they note that there is a strong dependence on the chosen radial surface density profile. 

 Planet occurrence rates around late-type stars can also inform us about the potential planet population around LTs. \citet{Dressing2015} examined the full \textit{Kepler} data set, and found an average planet occurrence rate of 2.5 $\pm$ 0.2 planets per M dwarf with radii between 1-4 $R_\oplus$. \citet{Ballard2016} analyzed \textit{Kepler} planet hosts using a dual-planet-population model, and found that about 50$\%$ of M dwarf planet hosts should have 5+ planets. Finally, \citet{He2017} produced the first upper-limits on planet occurrence rates around brown dwarfs, by searching for transits in lightcurves of 44 brown dwarfs observed with the \textit{Spitzer Space Telescope}. They found that the occurrence rate of planets within a 1.3 day orbit with radii between 0.75 and 3.25 R$_\oplus$ is less than 67 $\pm$ 1\%. 

The TRAPPIST-1 system also serves as strong observational evidence for the presence of a population of Earth-sized planets around LTs \citep{Gillon2017}. TRAPPIST-1 is an M8 main sequence star that has a radius of 0.121 $\pm$ 0.003 R$_\odot$, slightly larger than the characteristic radius of an LT \citep{VanGrootel2018}. Despite its small size, it hosts seven transiting, terrestrial exoplanets, all of which are about one Earth-radius in size. The coplanarity of these planets strongly suggests that they formed in a disk around TRAPPIST-1. A dedicated search for planets around LTs will help to address whether or not systems like this are common among the latest spectral types. 

Finally, detections of Jovian-mass objects have already been made around brown dwarfs at wide separations via gravitational microlensing and direct imaging. A directly-imaged 3-4 M$_{Jup}$ companion was discovered at a projected separation of 41 AU around a 25 M$_{Jup}$ brown dwarf \citep{Chauvin2004,Mamajek2005}, and a 0.75 M$_{Jup}$ planet was detected at a projected separation of 0.59 AU around a probable brown dwarf in the OGLE-2017-BLG-1522 system by \cite{Jung2018} through microlensing. These planets, with high companion-host mass ratios, likely formed via gravitational collapse \citep[e.g.,][]{Chauvin2005}. Lower-mass planets, formed by core accretion, have yet to be confidently detected around a brown dwarf. A Venus-mass planet around a brown dwarf was purportedly detected via gravitational microlensing by \citet{Udalski2015}, but subsequently has been contested \citep{Han2016}. 

This preponderance of evidence suggests that LTs host terrestrial planets, possibly in abundance. While remarkably different from familiar main-sequence stars, LTs present a number of advantages in the search for habitable, Earth-like planets that warrant a dedicated transit survey, as the field aims to characterize the first exo-Earth atmosphere in the coming decades.

In this work, we aim to devise strategies that maximize the number of detected planets for such a survey, by determining the detector type that produces the highest precision lightcurves, and optimizing the observing cadence.

This paper is outlined as follows: In Section \ref{sec:detector}, we investigate the performance of different detectors for performing a transit search of LT targets. In Section \ref{sec:simulation}, we describe simulated observations of LT planetary systems, which we use to develop an efficient observing strategy with a single telescope, and determine an optimal survey duration and telescope size. In Section \ref{sec:caveats}, we describe some limiting factors for searching for transiting planets around LTs. 

\section{Optimal Detector Setup for Earth-sized Planet Detection around LTs}
\label{sec:detector}

\begin{table*}[]
    \centering
    \begin{tabular}{|lrrrrrrr|}
        \hline
         Band& $\lambda_0$ ($\mu$m) & $\Delta\lambda$ ($\mu$m) & $N_R$ ($\frac{e^-}{pix}$) &  $\dot{N_D}$ ($\frac{e^-}{pix\cdot s}$) & $\dot{N_S}$ ($\frac{e^-}{pix\cdot s}$) & p ($\frac{''}{pix}$) & Size (pix)\\
        \hline
         \textit{z'} low dark current$^a$  &     0.9 & 0.12 & 1 & 5e-4 & 11 & 0.29 & 1024$\times$1024\\
         \textit{J} high dark current$^b$ &     1.2 & 0.16  & 87 & 163 & 47  & 0.87 & 640$\times$512 \\
         \textit{H} high dark current$^b$ &     1.7 & 0.25  & 87 & 163 & 247 &  0.87 & 640$\times$512 \\
         \textit{J} low dark current$^c$ &      1.2 & 0.16  & 19 & 23 & 22 & 0.59 & 1024$\times$1024 \\
         \textit{H} low  dark current$^c$ &     1.7 & 0.25  & 19 & 23 & 114 & 0.59 & 1024$\times$1024 \\
         \textit{K$_s$} low dark current$^c$ &  2.2 & 0.26  & 19 & 23 & 176 & 0.59 & 1024$\times$1024 \\
         \hline
    \end{tabular}
    \begin{flushleft}
        \footnotesize{$^a$\citet{Harding2016}};
        \footnotesize{$^b$\citet{Sullivan2014}};
        \footnotesize{$^c$\citet{Clemens2007}}
        
    \end{flushleft}
    
    \caption{Properties of the various bands used to simulate photometry. The central wavelength of each filter is given by $\lambda_0$, and $\Delta\lambda$ is the filter width around $\lambda_0$. The read noise associated with each detector is given by $N_R$, $\dot{N_D}$ is the dark current per pixel, and $\dot{N_S}$ is the rate of photoelectrons from the sky per pixel, estimated from sky brightnesses measured with the Perkins Telescope. The detector plate scale in arcsec pix$^{-1}$ is given by $p$. Finally, the detector size is given in pixels. The total throughput for each system was assumed to be 50\%.}
    \label{tab:instruments}
\end{table*}

To date, LTs have not been subject to a thorough search for transiting exoplanets. This is in part due to the fact that LTs are brightest in the near-infrared (NIR, $\sim$1.0-2.5 $\mu$m), where the performance of infrared array detectors has historically been worse than that of optical CCD detectors. In this section, we investigated the relative efficiency of performing a search for transiting exoplanets around LTs using simulated red-optical CCD and NIR array detectors.



\subsection{Detector Properties}
\label{sec:detectors}
We simulated photometry in six different bands, the properties of which are given in Table \ref{tab:instruments}. These properties were chosen to be broadly representative of modern detectors in use on research telescopes. We simulated a red-optical \textit{z'}-band detector, using the measured read noise and dark current from \textit{CHIMERA}, an optical imaging instrument on the 200-inch Hale Telescope \citep[][]{Harding2016}. We simulated high-dark-current NIR \textit{J}- and \textit{H}-bands, representative of economical InGaAs detectors, with properties taken from \citet{Sullivan2014}. Finally, we simulated low-dark-current NIR \textit{J}-, \textit{H}-, and \textit{K$_s$}-bands, with properties taken from \textit{Mimir}, a NIR polarimeter and imager in use on the 1.8-m Perkins Telescope \citep[][]{Clemens2007}. For each detector, we estimated the number of electrons per pixel per second resulting from sky brightnesses measured on the Perkins Telescope during bright time: 20.2 mag arcsec$^{-2}$ in \textit{z'}, 20.1 mag arcsec$^{-2}$ in \textit{J}, 18.0 mag arcsec$^{-2}$ in \textit{H}, and 16.8 mag arcsec$^{-2}$ in \textit{K$_s$}. We used the Perkins Telescope site as a reasonable continental observatory site where a transit search could be conducted. The total throughput for each detector was taken to be 50\%. In reality, the net throughput will not be equal for all systems, and the value used in this analysis only gives a sense of the relative performance of the different detectors. The plate scale and format of each detector, used for calculating the field of view (FOV), is also listed in Table \ref{tab:instruments}.


\subsection{Target Sample and FOV Limitations}
We assembled a sample of LT targets for which we simulated photometry, using the detectors described above. To create this sample, we selected all targets with measured \textit{z'}, \textit{J}, \textit{H}, and \textit{K$_s$} magnitudes from an online repository of ultra-cool dwarfs \citep[][and discovery references therein]{Gagne2014}\footnote{https://jgagneastro.wordpress.com/list-of-ultracool-dwarfs/}. All objects in this list have a measured spectral type of L0 or later, and 132 targets had all four magnitude measurements.

Before simulating photometry for these targets, we considered the number of suitable reference stars available within each detector's FOV. Bright reference stars are necessary for removing changes in brightness that are common to all objects in a field, generally due to changes in observing conditions. However, if an object does not have reference stars that are as bright or brighter than it within the FOV of the detector, correcting variability in a target lightcurve by dividing by a reference lightcurve can increase the scatter in the target lightcurve. We therefore treated the effects of incorporating reference lightcurves explicitly, by identifying suitable references near each target, simulating their summed lightcurve, and dividing the target lightcurve by the summed reference lightcurve.

First, we calculated the FOV for each detector, using the plate scales and detector sizes listed in Table \ref{tab:instruments}. For the high-dark-current NIR detectors, with unequal height and width, we used the smaller of the two dimensions. We searched for reference stars within the FOV of each detector for every target in the sample. We queried the 2MASS Point Source Catalog \citep{Cutri2003} for nearby reference stars for the $J-$, $H-$, and $K_s$-band detectors, and SDSS Data Release 12 \citep{Alam2015} for the $z'$ detector. We made a reference lightcurve for each target using reference stars that were as bright or brighter than the target.

Following \citet{Nutzman2008}, we required the reference lightcurve to have at least 10 times as many counts as the target lightcurve. We found that all targets met this requirement for all of the simulated detectors, except for two targets in high-dark-current \textit{J}-band and six targets in high-dark-current \textit{H}-band (1.5\% and 4.5\% of the sample, respectively). Because of their intrinsic faintness, there are generally several reference stars within the FOV of the simulated detectors that are brighter than the target LT. For this reason, the noise introduced by reference lightcurves is low, and detector FOV is not a limiting factor in producing differential photometry, provided one has access to a detector with an FOV of several arcminutes.

\subsection{Photometry}
We simulated photometry for targets that had a sufficient number of nearby reference stars using each simulated detector. We assumed a 2-m telescope with a 5\% central blockage, 30-second exposure times and 1".5 FWHM seeing. The SNR of each lightcurve was calculated using the CCD equation: 
\begin{equation}
    \frac{S}{N} = \frac{\dot{N_*}t}{\sqrt{\dot{N_*}t+n_{pix}(\dot{N_S}t+\dot{N_D}t+N_R^2)}}
\end{equation}
where $\dot{N_*}$ is the rate of photoelectrons per second from the target, $\dot{N_S}$ and $\dot{N_D}$ are the rates of photoelectrons per pixel per second from sky background and dark current, respectively, $N_R$ is the number of photoelectrons from read noise per pixel, $t$ is the exposure time, and $n_{pix}$ is the area of the photometric aperture in pixels. For each target, we determined $n_{pix}$ by calculating the SNR for different aperture radii, ranging from 0.1 to 2 times the seeing FWHM in steps of 0.05. The radius that resulted in the highest SNR for the target was chosen for performing photometry. 

$\dot{N_*}$ was determined using:
\begin{equation}
    \dot{N_*} = \Phi_\lambda(0) \times \Delta\lambda \times 10^{-m/2.5} \times A \times f_{through} \times f_{Gauss}
\end{equation}
where $\Phi_\lambda(0)$ is the number of photoelectrons per second per area of aperture per unit wavelength for a 0-magnitude star, as calculated for the 2MASS \textit{JHK$_s$}-bands \citep{Cohen2003} and SDSS \textit{z'}-band \citep{Fukugita1996} (assuming a gain of 1), $\Delta\lambda$ is the filter width for the respective band, as given in Table \ref{tab:instruments}, $m$ is the target magnitude in the respective band, $A$ is the telescope aperture area, and $f_{through}$ is the fraction of photoelectrons that are observed after accounting for throughput losses; here, we assumed $f_{through} = 0.5$ for every detector. Finally, $f_{Gauss}$ is the fraction of stellar light that falls within the optimal aperture radius for each target, assuming Gaussian PSFs with FWHM = 1".5.

By normalizing the signal to one and assuming Gaussian errors, the standard deviation of the lightcurves is found by simply inverting the SNR equation. We also made a normalized master reference lightcurve for each target in each band, consisting of the summed counts from suitable reference stars near each target. The target lightcurves were then divided by these reference lightcurves to produce a final lightcurve for each target for all six of our simulated detectors. 

\begin{figure*}
    \centering
    \includegraphics[width=\textwidth]{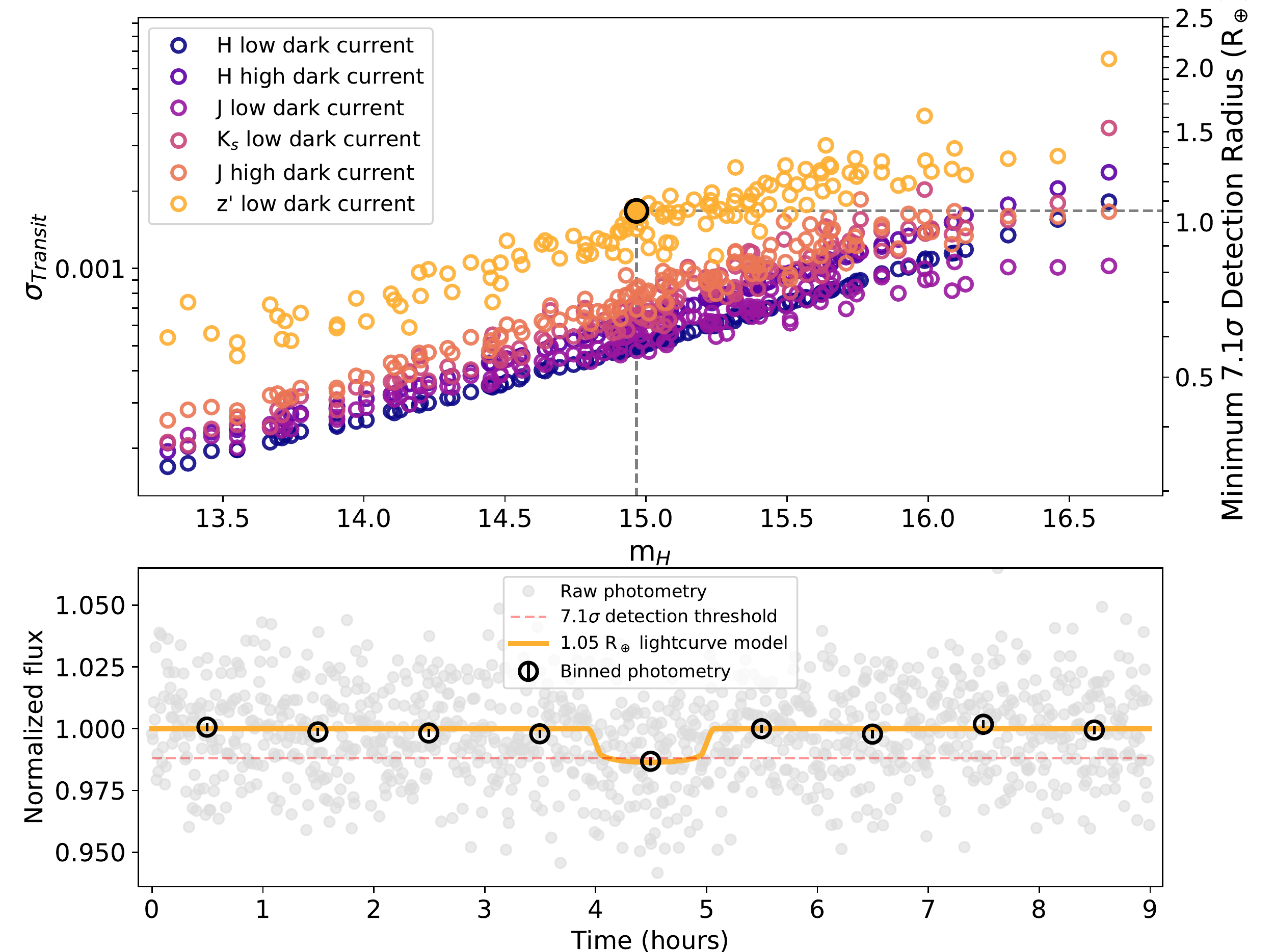}
    \caption{\textit{Top}: The photometric error on the timescale of one hour vs. the magnitude of each target in \textit{H}-band for the six different simulated detectors. These results are based on a 2-m telescope using a 30-second exposure time, and every detector is assumed to have a 50\% net throughput. The plot is also labeled with the minimum planetary radius that could be detected at 7.1$\sigma$ in each lightcurve, assuming transits in front of a typical 0.88-R$_{Jup}$ LT \citep{Burgasser2001}. The scatter in each series is due to the different reference stars available for each target. \textit{Bottom}: A simulated lightcurve for the highlighted \textit{z'}-band point in the top panel. The 30-second exposures (grey points) have been binned over one hour (black circles with error bars). The 7.1$\sigma$ detection threshold of the data is indicated with a red dashed line, and a one hour transit of a 1.05-R$_\oplus$ planet has been injected into the data.}
    \label{fig:efficiency}
\end{figure*}

\subsection{Lightcurve Scatter for Different Detectors}
\label{sec:lc_scatter}
We then binned the final lightcurves over a time of one hour (i.e. 120 exposures), the timescale of a transit of a habitable planet in front of an LT, assuming an inclination of 90$^\circ$, a period of 1.65 days, and an a/R$_*$ of 17. In the top panel of Figure \ref{fig:efficiency}, we plot the standard deviation of lightcurves on the one-hour timescale for all detector setups and all targets in the sample. As this plot shows, the NIR detectors all gave better precision than the red-optical detector, owing to the faintness of LT targets in \textit{z'}-band. Despite the higher sky background, the low-dark-current \textit{H}-band detector produced the lowest-scatter lightcurves on average, because LTs are typically brightest at \textit{H}-band wavelengths, and because \textit{H}-band is fairly wide. On average, these detectors produced lightcurves with standard deviations a factor of 2.7 times smaller than those produced with the \textit{z'} detector. 

The higher scatter of the \textit{z'}-band photometry translates to a lower detection efficiency  of Earth- and super-Earth-sized planets. The right-hand y-axis of Figure \ref{fig:efficiency} is labeled with ``minimum detection radii": the planetary radii that would produce a 7.1$\sigma$ transit depth around a typical 0.88-R$_{Jup}$ LT \citep{Burgasser2001} if the transit were fully captured in one bin. We chose this detection threshold both because of its use in large-scale transit surveys like \textit{Kepler} \citep{Jenkins2002} and \textit{TESS} \citep{Sullivan2015} (note that these two studies arrived at the same threshold by coincidence), and because it virtually guarantees zero non-astrophysical false positives over the course of the surveys simulated in Section \ref{sec:simulation}. We discuss the choice of this threshold in more detail in Section \ref{sec:overview}.

In the bottom panel of Figure \ref{fig:efficiency} we show the lightcurve of the highlighted \textit{z'} point from the top panel, for the target with the median magnitude in the sample (m$_H \sim$ 15). In this lightcurve, we injected the transit of a planet with the minimum detection radius, 1.05 $R_\oplus$, using the \texttt{BATMAN} software package \citep{Kreidberg2015}. The in-transit point just crosses the 7.1$\sigma$ detection threshold for the lightcurve, indicating that for 50\% of the sample, \textit{z'} observations would be insensitive to the detection of planets smaller than about 1.05 $R_\oplus$. For low-dark-current \textit{H}-band observations, however, this value is about 0.60 R$_\oplus$.




It should be emphasized that this analysis was performed neglecting sources of systematic noise \citep[see, e.g.,][]{Croll2015}, which would serve to increase the minimum detectable planet size for all of our simulated detectors. One such source of systematic noise that is particular to very red objects is differential extinction. LTs are generally redder than nearby reference stars, and the spectral energy distributions (SEDs) of the target and references are attenuated differently by the Earth's atmosphere, leading to systematic differences in relative photometry \citep[e.g.,][]{Bailer-Jones2003,Berta2012}. \citet{Blake2008} found that this effect can be large in \textit{J}-band, with short-duration changes on order of 1\%, similar to the transit depths anticipated for Earth-sized planets around LTs. However, it is less significant in \textit{H}- and \textit{K$_s$}-bands (\textless 0.3\%), because these bands largely avoid significant atmospheric absorption features.

As a result of this analysis, and considering the effects of differential extinction, we find that detectors with \textit{H}-band imaging capabilities offer the best performance for searching for Earth-sized planets around LTs. Both low- and high-dark-current variants offer similar performance, enabling the possibility of using more economical InGaAs detectors for performing such a search. Low-dark-current \textit{z'} detectors offer the worst performance, but are sensitive to the detection of Earth-sized planets for about 50\% of the sample.

\subsection{Lightcurve Scatter as a Function of Telescope Size} \label{sec:tel_scatter}
\begin{figure}
    \centering
    \includegraphics[width=\columnwidth]{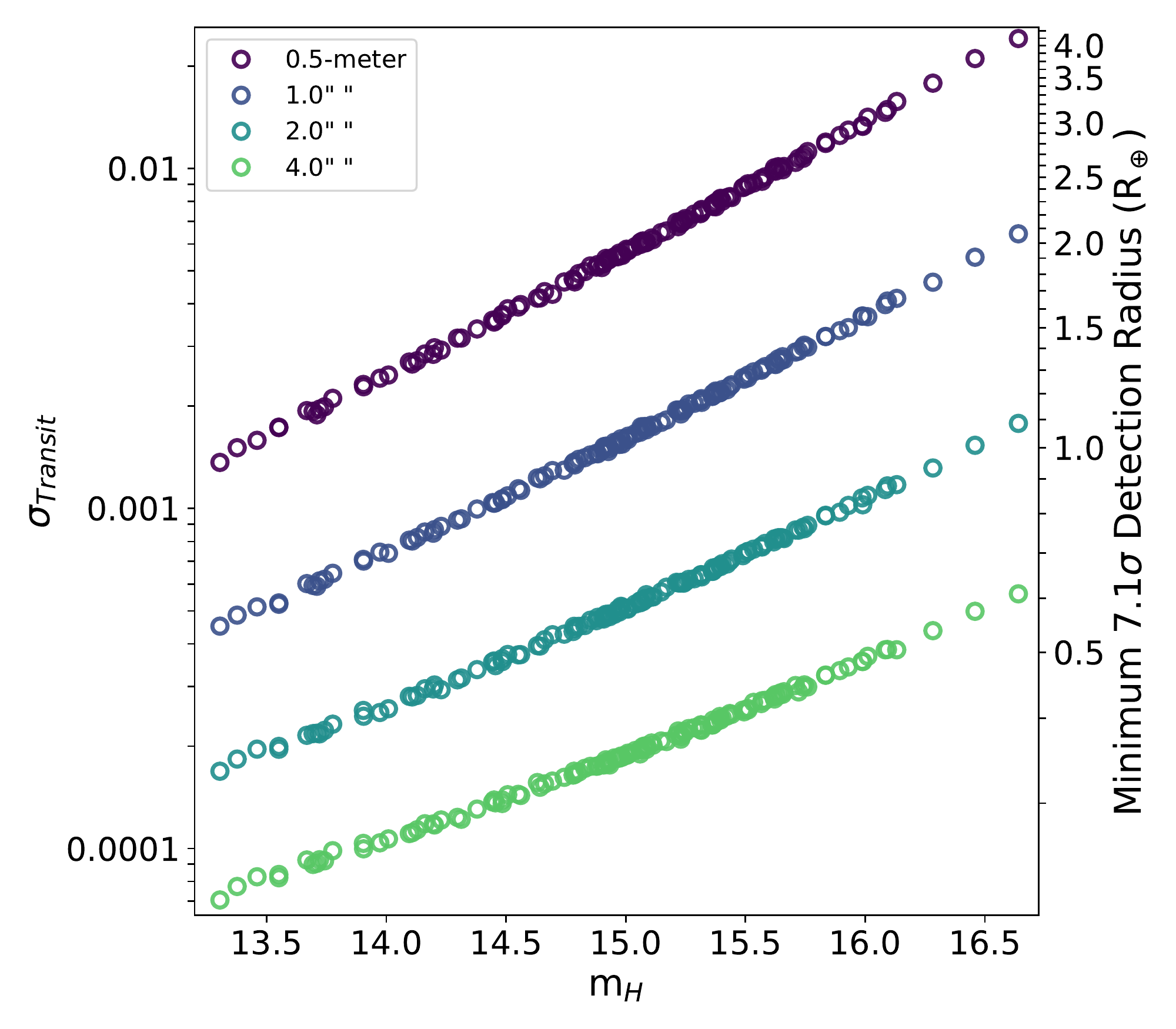}
    \caption{Standard deviation of simulated lightcurves on the timescale of 1-hour versus target magnitude in \textit{H}-band for different sized telescopes. Photometry was simulated using the low-dark-current \textit{H}-band detector described in Sec. \ref{sec:detectors}. As the 7.1$\sigma$ detection radii indicate, telescopes smaller than 1-meter are insensitive to the detection of Earth-sized planets for a majority of LT targets.}
    \label{fig:detection_efficiency}
\end{figure}

We performed a similar analysis for a variety of telescope sizes, the results of which we show in Figure \ref{fig:detection_efficiency}. We measured the scatter on 1-hour timescales in photometry from our simulated low-dark-current \textit{H}-band detector for the same sample as above, varying the telescope diameter between 0.5, 1.0, 2.0, and 4.0-m. As the figure demonstrates, telescopes that are smaller than 1-m  will be insensitive to the detection of Earth-radius planets for a majority of LT targets, even using the optimal detector setup found in the previous section. Telescopes that are 2-m and larger are sensitive to the detection of Earth-sized planets over the entire sample. 

In terms of minimum detection radius, increasing the telescope size has diminishing returns for the purpose of detecting Earth-like planets. Larger telescopes, however, would enable the detection of sub-Earth-sized planets. The median minimum detection radius is 1.93 R$_\oplus$ for a 0.5-m telescope, 1.03 R$_\oplus$ for a 1-m, 0.57 $R_\oplus$ for a 2-m, and 0.35 R$_\oplus$ for a 4-m. Given these results, telescopes with aperture diameters of 1- to 2-m should be sufficient for conducting surveys for transiting Earth-analogs around samples of LTs.


\section{Simulated Survey of LTs} \label{sec:simulation}

In this section, we investigated the potential for performing a search for transiting exoplanets around a sample of LTs using a single telescope with a low-dark-current NIR imager, the optimal detector determined in the previous section. We simulated observations of planetary systems around a sample of LTs over durations of one to five years, with telescope sizes from 0.5 to 4.0 meters. We assessed the likelihood of detecting at least one planet for a variety of observational strategies and used this to develop an optimal survey design.

\subsection{Survey Simulation Overview}
\label{sec:overview}

\begin{figure*}
    \centering
    \includegraphics[width=\textwidth]{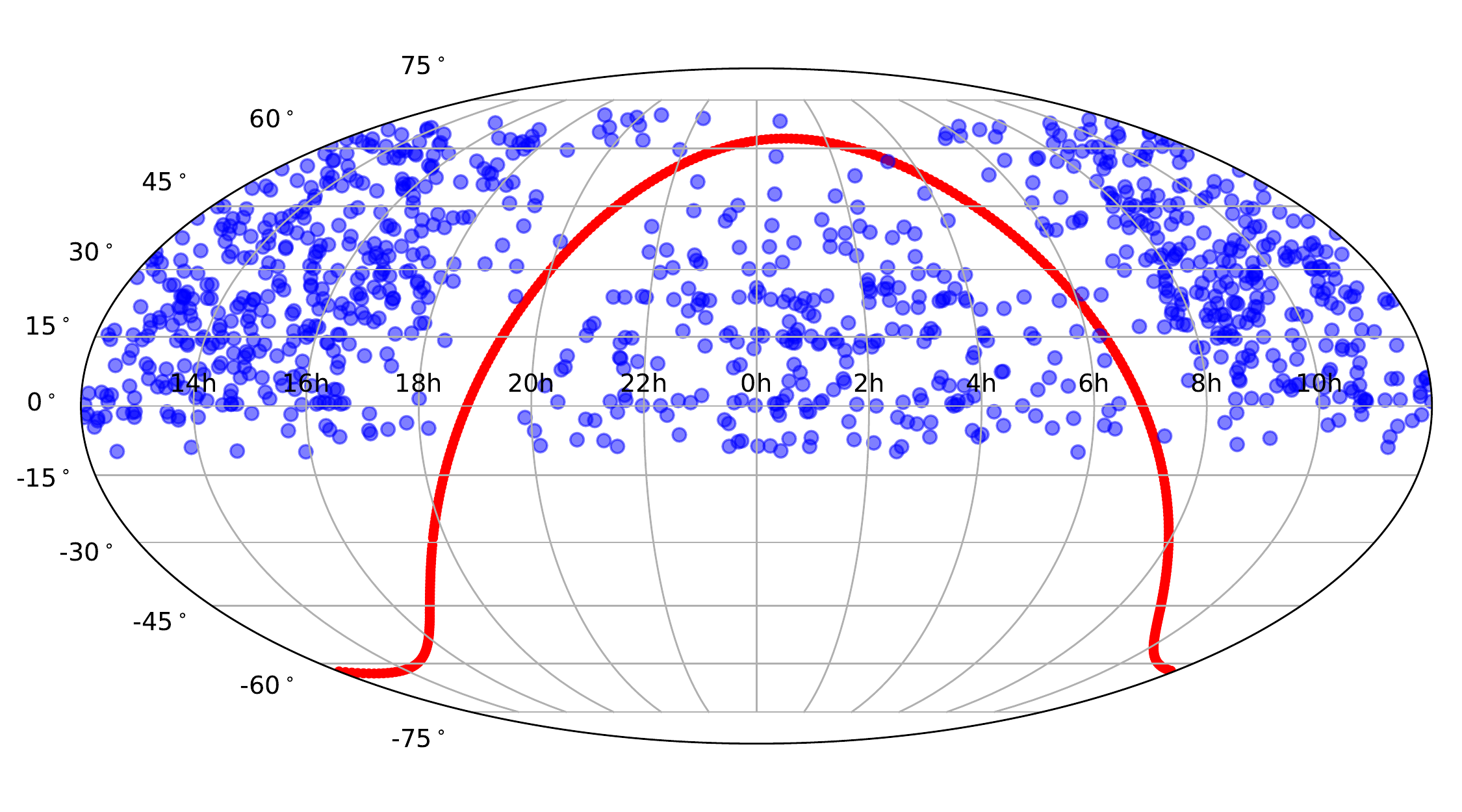}
    \caption{Locations on the sky for the 998 LT targets within the declination limits of the Perkins Telescope. The galactic plane, which inhibits the discovery of faint LTs, is shown in red.}
    \label{fig:sky_map}
\end{figure*}

\begin{figure}
    \centering
    \includegraphics[width=\columnwidth]{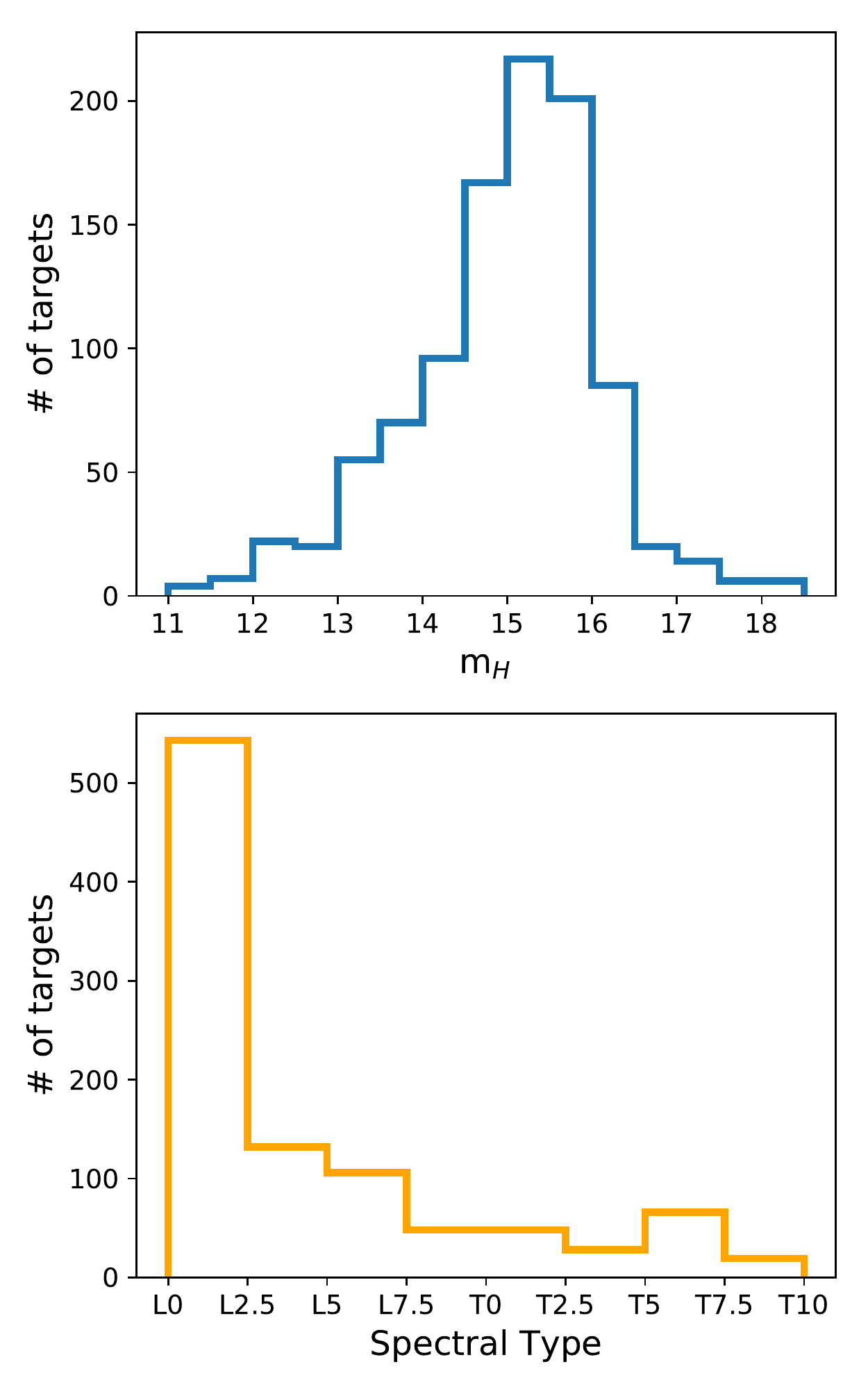}
    \caption{\textit{Top}: Histogram of the \textit{H}-band magnitudes for the 998 targets in our sample. The median magnitude is 15.1. \textit{Bottom}: Histogram of the spectral types of targets in the sample.}
    \label{fig:hists}
\end{figure}

 We began by assembling a sample of LTs that are observable from a particular latitude on Earth. We chose this latitude to be that of the 1.8-m Perkins Telescope in Anderson Mesa, AZ (+35$^\circ$ 05' 48.6"), and selected targets within the telescope's declination limits (-10$^\circ < \delta < $ 70$^\circ$). These targets were chosen from the same database used in Section \ref{sec:detector}, and all have measured NIR or optical spectral types of L0 and later. We found 998 LT targets that are observable from the chosen latitude, and show a sky-map of these targets in Figure \ref{fig:sky_map}. Histograms of target magnitudes and spectral types can be found in Figure \ref{fig:hists}.  Seven percent of the sample resides in the L/T transition \citep[i.e., spectral types L9 to T3.5,][]{Radigan2014a}, where enhanced photometric variability has been observed due to the dissipation of cloud features in LT atmospheres \citep[e.g.,][]{Radigan2014b}. 
 
 Forty percent of the sample is spectral type L4 or later (398 targets), where objects are brown dwarfs or planetary mass objects. This is a higher fraction of substellar objects than is targeted by the SPECULOOS survey, for which $\sim$10\% of their 1136-target sample is estimated to be brown dwarfs \citep{Delrez2018}. This is largely because SPECULOOS operates at red-optical wavelengths, and is hence more sensitive to detecting planets around slightly earlier spectral types ($\sim$86.7\% of the SPECULOOS sample is spectral type M7-M9).

We applied a different \textit{H}-band magnitude cutoff to the sample of targets observable from Perkins as a function of telescope size, using the results of Section \ref{sec:tel_scatter} (see Figure \ref{fig:detection_efficiency}). We chose this cutoff to be the magnitude beyond which a given telescope would be insensitive to the detection of 2.0 R$_\oplus$ planets, which is near the maximum size of terrestrial silicate planets \citep[e.g.,][]{Seager2007}. We determined these cutoffs via linear fits to the series plotted in Figure \ref{fig:detection_efficiency}. The selected cutoffs for each telescope size are given in Table \ref{tab:samples}, along with the number of potentially observable targets below the cutoff. 

\begin{table}[]
    \centering
    \begin{tabular}{|c|c|c|}
    \hline
       Diameter (m) & m$_H$ cutoff & Number of targets \\
       \hline
        0.5 & 14.9 & 196 \\
        1.0 & 16.5 & 764 \\
        2.0 & - & 998 \\
        4.0 &  - & 998 \\
        \hline
    \end{tabular}
    \caption{\textit{H}-band magnitude cutoffs and number of potentially observable targets for each telescope size for the simulation described in Section \ref{sec:simulation}.}
    \label{tab:samples}
\end{table}

To construct lists of observing nights, we assumed 10 nights of observation per month during bright time, for one to five years. For a given combination of target sample (see Table \ref{tab:samples}) and survey duration, we then scheduled the observation of targets. On each date, we selected targets for observation out of groups which transited the meridian within two hours of local midnight. The brightest observable targets were prioritized in the scheduling. If no new targets were available for scheduling, we scheduled previously-observed targets. 

We assigned effective temperatures to these targets using the $T_{eff}$-spectral-type relation for field M6-T9 dwarfs from \citet{Faherty2016}. We also assigned each object a random age, drawn from a uniform distribution between 0.0005 and 10 Gyr. We used these ages and temperatures to identify the closest matching evolutionary model point from \citet{Baraffe2015}, from which we took the corresponding mass and radius. The median mass of the sample was 78.6$^{+5.2}_{-5.2}$ M$_{Jup}$, and the median radius was 0.92$^{+0.07}_{-0.02}$ R$_{Jup}$, reflective of the sample's bias towards early L-dwarfs.

We then simulated planetary systems around these targets using measured M dwarf planet occurrence rates from \citet{Dressing2015}. 
For each target, we stepped through the radius-period grid from \citet{Dressing2015}, creating planets if a randomly drawn number was less than the occurrence rate in that grid space. We note that this is likely a conservative approach. The occurrence rate of short-period planets increases with later spectral type, with occurrence rates around M dwarfs being about twice as high as those for G dwarfs, and three times higher than those for F dwarfs \citep[][]{Mulders2015}. In addition, \citet{Hardegree2019} demonstrated that the occurrence rate of short-period planets increases \textit{within} the M spectral type, with M5's having rates roughly 3.5 times higher than M3's (although with large error bars). If these trends continue into the LT spectral types, the \citet{Dressing2015} M dwarf rates likely underestimate the population of planets around LTs. In addition, these rates are limited to planets with orbital periods with radii between 0.5 and 4.0 R$_\oplus$. In principle, planets with sizes outside of this range can exist around LTs, but we did not simulate them.

If simulated, a planet was assigned a random period and radius within its grid space, with its orbital separation determined using Kepler's third law and assuming zero eccentricity. We then assigned each system a random on-sky inclination, assuming equal likelihood of the orbital axis across 4$\pi$ steradians (i.e., equal probability over the area of a unit sphere). If multiple planets were created around one target, we assigned mutual inclinations with respect to the system inclination by drawing from a normal distribution with width 0.3$^\circ$. This value represents the 90\% confidence upper limit for the scatter in the mutual inclinations between the TRAPPIST-1 planets, as calculated by \citet{Luger2017}.

We used the geometry of each system (host size, planet size, planet orbital distance and system inclination) to determine which planets would transit from our line-of-sight. Doing this, we found that 0.8$^{+0.3}_{-0.3}\%$ of all targets have at least one transiting planet, on average. We finished creating the synthetic planet populations by assigning each planet a random orbital phase. We then calculated ephemerides using each planet's assigned period, and generated a list of all transit events over the course of the survey. 

We simulated photometry using the properties of the low-dark-current \textit{H}-band detector listed in Table \ref{tab:instruments}, but we changed the total throughput to 41.5\% to more closely match the measured value for the NIR \textit{Mimir} instrument \citep{Clemens2007}. If a transit was observed, we examined the target's simulated lightcurve. We injected transits into the lightcurves using the \texttt{BATMAN} software package \citep{Kreidberg2015}. We assumed a quadratic limb darkening law, with parameters appropriate for a 2000 K target in \textit{H}-band taken from the tables of \citet{Claret2011}. We found u$_1$ = 0.3455, and u$_2$ = 0.4918. 

The lightcurves with injected transits were then binned in time. Data were binned such that each ``block" of data was collapsed to one point (see Section \ref{sec:optimization} and Figure \ref{fig:sim_detection}). We then checked if each lightcurve contained points that exceeded a chosen detection threshold, which we took to be 7.1$\sigma$. This is a conservative value, as it virtually guarantees zero non-astrophysical false positives over the duration of the simulated surveys, which we illustrate through the following numerical example. Assuming 30-second exposure times, a two-minute overhead for slewing between targets, and 10-hour average night lengths, we expect a typical survey to take roughly 1000 exposures per night. For a 5-year survey with no weather losses, this results in 6e05 exposures. Assuming Gaussian uncorrelated noise, we expect a single exposure to be a 7.1$\sigma$ one-tailed outlier roughly with a frequency of 1 in 1.6e12, a number vastly lower than the number of exposures. 
Because we are not explicitly modeling sources of systematic noise, which would results in significant outliers more frequently, and because of the historical significance of 7.1$\sigma$ in the \textit{Kepler} and \textit{TESS} surveys, we take 7.1$\sigma$ as the detection threshold for the simulation.

If a lightcurve contained at least one point that was decremented below 7.1$\sigma$ of the binned data, and a planet was truly present in the lightcurve, we considered that planet to be ``detected" \citep[i.e., similar to][]{Sullivan2015}; in reality, such a signal would trigger future follow-up observations to check whether the signal is periodic (see Section \ref{sec:caveats}). 

\subsection{Survey Optimization}

\label{sec:optimization}
Several different parameters, corresponding to different observing cadences, could affect the likelihood of success of a given survey. These include the number of nights spent observing a target/group of targets, the number of targets observed per night, and the amount of time spent observing an individual target before switching to the next. We explored grids of these parameters, recording the success rates for different combinations of telescope diameter, survey duration, and observational strategies over 2000 simulations. 

We allowed the number of nights per target group to vary from 1 to 20 days in intervals of 1 day. We maximized the planet yield when observing each target group for seven days. However, because we assumed 10-night observing runs, we elected to allot five nights for observing each group in all our simulations. This allowed us to schedule two full groups per run, without having to carry over observations of groups between different runs. We found that using five nights per group did not strongly impact the predicted yield of planets compared to seven nights.

\begin{figure}
    \centering
    \includegraphics[width=\columnwidth]{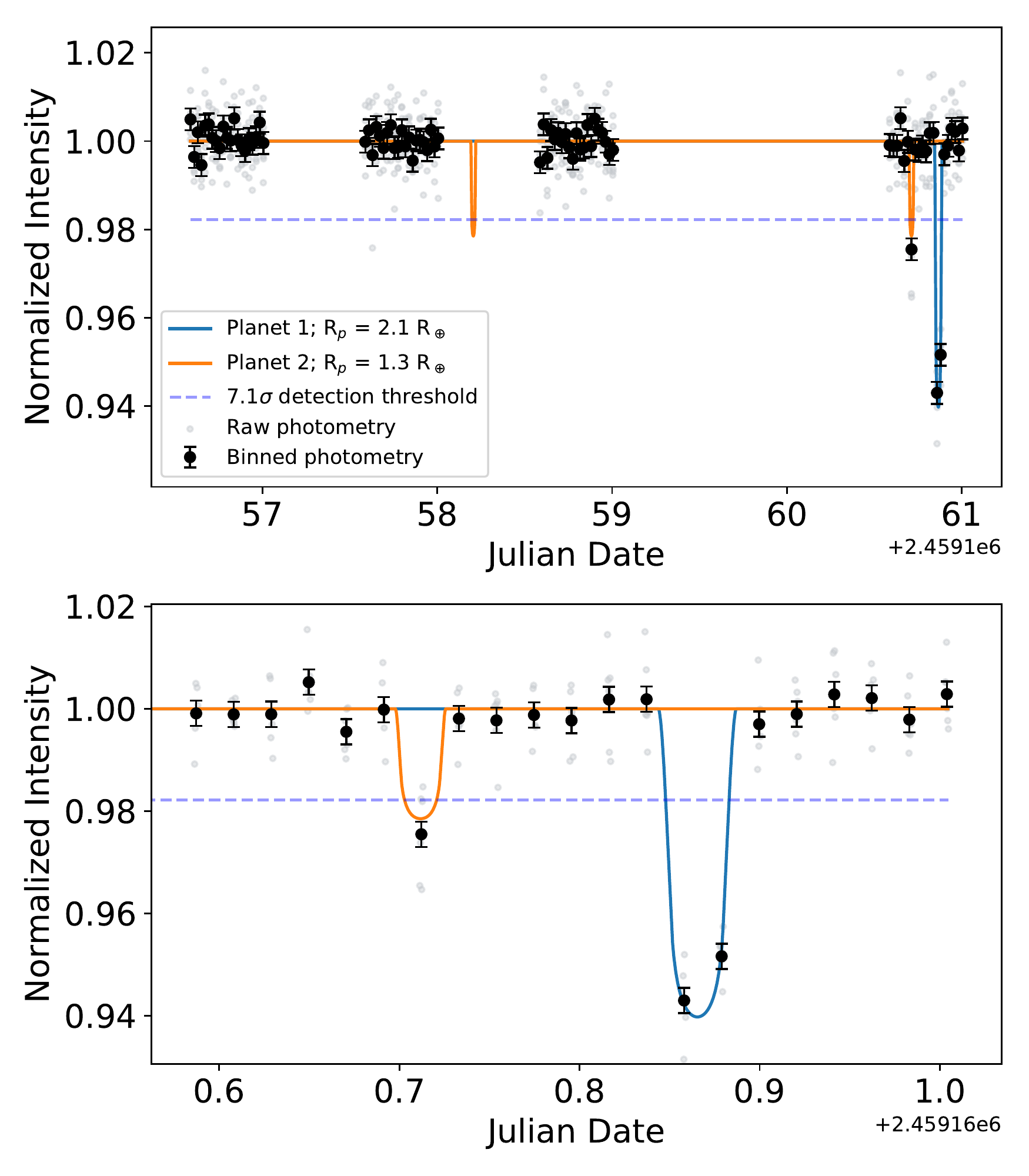}
    \caption{A simulated lightcurve for a target with two detected transiting planets. The target had a measured \textit{H}-band apparent magnitude of 14.9, a simulated radius of 0.86 R$_{Jup}$, and a simulated mass of 46.1 M$_{Jup}$. \textit{Top}: The full five-night lightcurve for the object, with one night lost randomly due to weather. One transit occurred during the day, when no data were being taken. Photometry was simulated using a 2-m telescope in low-dark-current \textit{H}-band, a 30-second exposure time, and spending five minutes on the target before slewing. Black points with error bars show data binned over three minute intervals (five minutes minus two minutes of overhead). A dashed blue line indicates the 7.1$\sigma$ detection threshold for the lightcurve. \textit{Bottom}: Night five only, which shows the transits of 1.3 and 2.1 R$_\oplus$ planets. Gaps in the data correspond to times when other targets were being observed. }
    \label{fig:sim_detection}
\end{figure}

We found that staring at single targets each night vastly limits the feasibility of performing this search with a single telescope. The likelihood of detecting planets can be improved by ``dithering" between $n$ targets per night, which effectively increases the length of time that can be spent on an individual target by a factor of $n$. Although this strategy will not return entire transit events, the presence of transits can be revealed by individual blocks of data that are significantly displaced below the average level. This is demonstrated in Figure \ref{fig:sim_detection}, where we show a typical detection lightcurve for one of the targets. MEarth also employs a dithering strategy \citep{Nutzman2008}, but we emphasize that that survey uses two robotic arrays consisting of eight telescopes each, whereas this simulation is being performed assuming a single telescope. We varied the number of targets observed each night between two and seven. 

We also varied the time spent observing one target before dithering to the next, with values of 2.5, 5, 10, and 20 minutes. We assume 2 minutes of overhead built into each value, to simulate time that would be wasted slewing and acquiring the new target each switch. For example, targets observed with the five-minute cadence are given three minutes of simulated observations, and two minutes of empty time. Assuming a uniform 30-second exposure time, as we did, cadences shorter than 2.5 minutes are impossible. This dithering strategy results in ``blocks" of data, during which each target is briefly observed. We bin over these blocks, and use the binned data to check if the 7.1$\sigma$ detection threshold is crossed (see Figure \ref{fig:sim_detection}).

\begin{figure*}
    \centering
    \includegraphics[width=\textwidth]{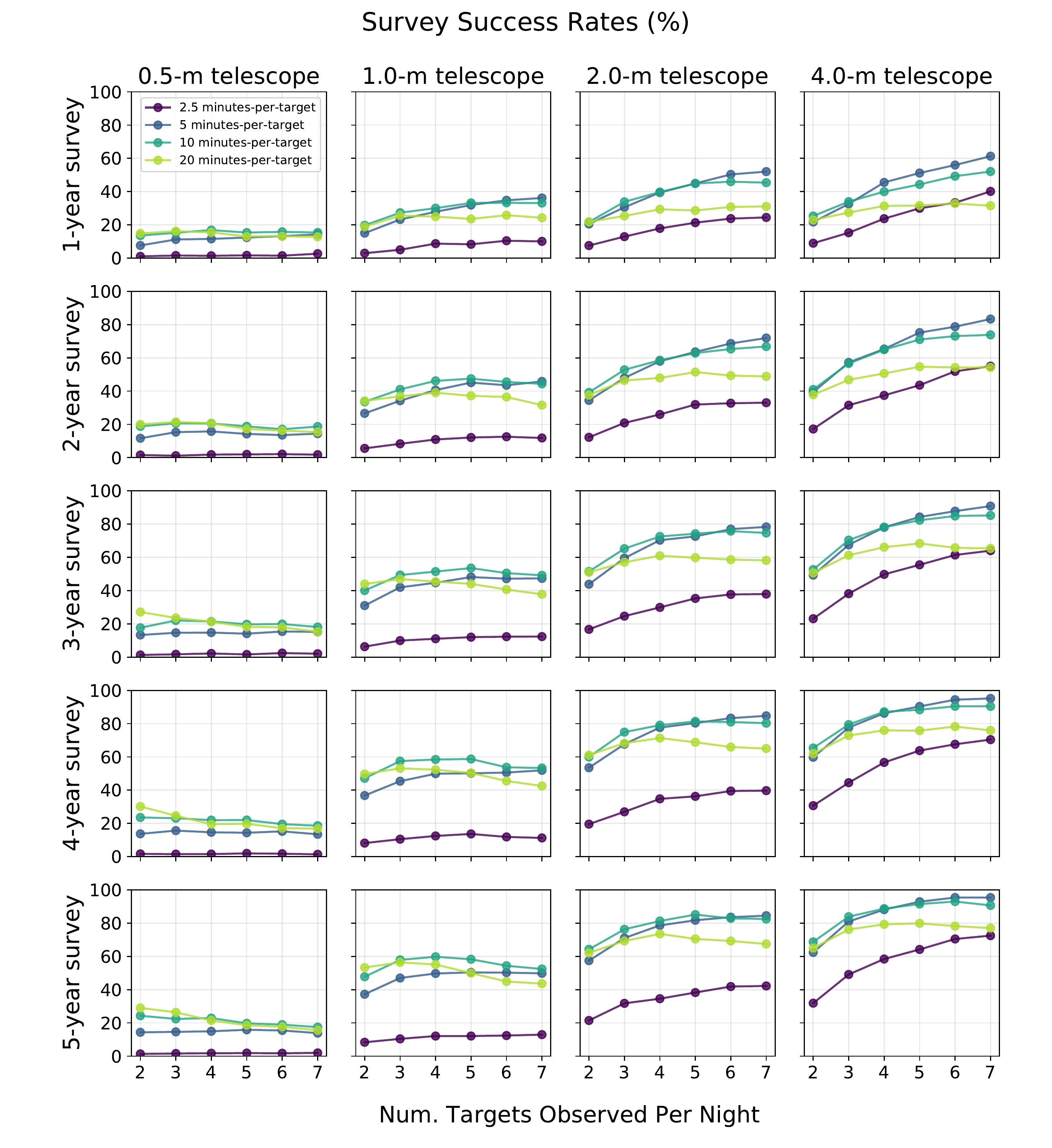}
    \caption{Success rates (defined as the fraction of surveys in which at least one planet was detected at greater than 7.1$\sigma$) for a variety of different survey configurations. Telescope diameter increases to the right, and survey duration increases to the bottom. In each panel, we plot survey success (in percent) as a function of the number of targets observed each night for the given combination of survey duration and telescope size. The four series shown in each panel correspond to strategies where different amounts of time were spent observing each individual target before slewing to the next.}
    \label{fig:optimization}
\end{figure*}

\begin{figure*}
    \centering
    \includegraphics[width=\textwidth]{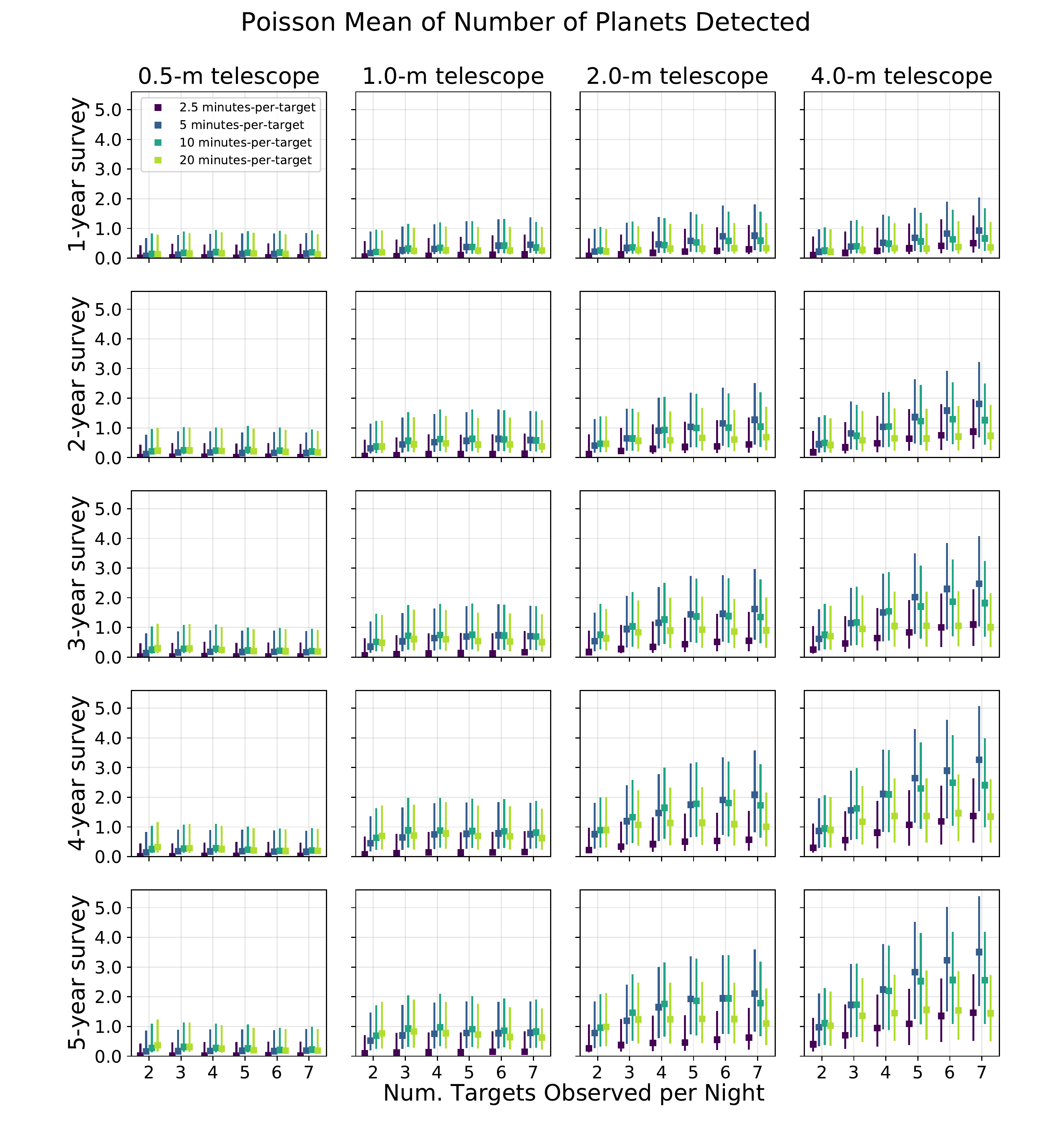}
    \caption{Poisson mean of number of detected planets for the same survey configurations shown in Figure \ref{fig:optimization}. We recorded the number of planets detected for each configuration over 2000 simulations, then fit a Poisson curve to this distribution. The mean of each Poisson fit is shown, along with the 16th and 84th percentile range of the distribution. Points are offset horizontally for clarity.}
    \label{fig:poisson}
\end{figure*}

We ran the simulation 2000 times for each combination of telescope size, survey duration, number of targets per night, and number of minutes per target ($4 \times 5 \times 6 \times 4 = 480$ combinations). For each, we recorded the survey's success rate, defined to be the fraction of simulations in which the given combination detected at least one planet at greater than 7.1$\sigma$ confidence. We show the results of this analysis in Figure \ref{fig:optimization}, where we plot success rate as a function of the number of targets observed per night for every combination.

Several trends are apparent in Figure \ref{fig:optimization}. First, longer duration surveys with larger telescopes are generally more successful, as expected. Longer surveys look at more targets than shorter surveys, and larger telescopes are more sensitive to the detection of planets than smaller telescopes (see Figure \ref{fig:detection_efficiency}). 

However, Figure \ref{fig:optimization} also shows diminishing returns as the survey duration is increased. By the end of a four-year survey (here, 480 observing nights before accounting for weather), most of the  brightest observable targets in each sample had been observed for at least five nights. With the sample mostly explored, any planets likely to be detected already had been, and searching for longer did not translate to a substantial increase in survey success. A four-year survey performed in this manner is more efficient than a five-year one. 

Figure \ref{fig:optimization} also shows diminishing returns in survey success with increasing telescope aperture size. This is because past a certain telescope size, a survey is essentially sensitive to the detection of all simulated planets that may appear in the photometry. This was previously demonstrated in Figure \ref{fig:detection_efficiency}, in which a 2-m telescope was sensitive to the detection of all Earth-sized planets over the sample of LTs. While a 4-m class telescope gives higher probability of success (again, detecting at least one planet), the increase is marginal. 

Finally, trends can be seen within each panel, due to differences between the number of targets observed each night and the amount of time spent observing targets. More successful strategies observe between five and seven targets each night, and spend between five and ten minutes observing each target before switching to the next. The 20-minute cadence becomes less effective as more targets are observed each night, because transits can occur in between successive observations of an individual target. The 2.5-minute cadence generally performs the poorest, although it starts to become viable as more targets are observed each night.

We also show the mean number of planets detected using each configuration in Figure \ref{fig:poisson}. We recorded the number of detections in each iteration of the simulation for every survey strategy, constructed histograms of these detections, and fit Poisson curves to these distributions. In Figure \ref{fig:poisson}, we show the mean of the fitted Poisson functions, with error bars marking the 16th and 84th percentiles of the distributions. This figure displays many of the same trends visible in Figure \ref{fig:optimization}, but translated to actual number of planet detections. Short-duration surveys conducted with small telescopes detect fewer than one planet on average, while long-duration surveys with large telescopes can detect upwards of two. This figure also shows that performing a survey with a 4-m telescope leads to the detection of a greater number of planets compared to a 2-m. 

Based on these results, we find that an optimal search for transiting planets around LTs uses a 2-m class telescope with a NIR instrument, a survey duration of 3-4 years (assuming 120 nights per year with 30$\%$ random weather losses), observes 5-7 targets each night, and slews between individual targets every 5-10 minutes. We find that surveys conducted in this way have a 73-85\% chance of making at least one 7.1$\sigma$ planet detection over the course of the survey (see Figure \ref{fig:optimization}), and have a 72\% chance of detecting between 1 and 3 planets, on average  (see Figure \ref{fig:poisson}). However, these results rely heavily on the assumed planet occurrence rates, and as noted previously,  the \textit{Kepler} M dwarf rates from \citet{Dressing2015} may underestimate the true planet population around LTs. Doubling these occurrence rates, following the results of \citet{Hardegree2019} for mid-type M dwarfs, roughly doubles the number of planets detected in the simulated surveys.

\subsection{Typical Detections from an Optimal Survey} \label{results}
 To determine what a typical planet detection would look like, we recorded the properties of planets recovered over 2000 simulations using a four-year survey with a 2-m telescope, observing six targets per night, and spending five minutes per target before slewing.

\begin{figure*}
    \centering
    \includegraphics[width=\textwidth]{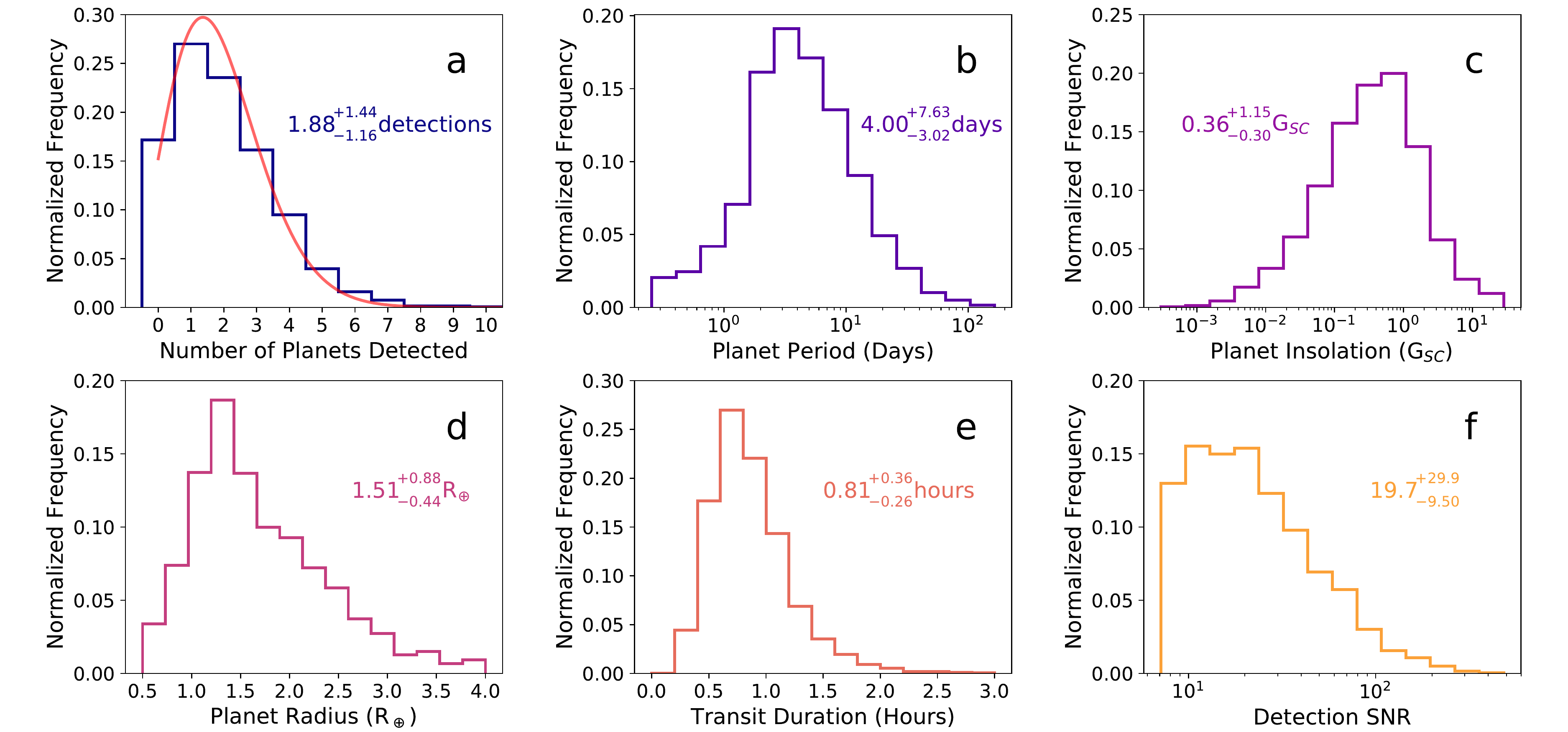}
    \caption{Normalized distributions of detected planet parameters for 2000 simulations of a four-year survey using a 2-m telescope, observing six targets per night, and spending five minutes per target before slewing. In panel \textit{a}, we report the median value of a Poisson fit to the distribution, along with the 16h and 84th percentile range of the Poisson curve. In panels \textit{b}, \textit{c}, \textit{d}, \textit{e}, and \textit{f}, we report the median value for the  parameter in question, along with uncertainties based on the 16th and 84th percentiles of the distribution.  \textit{Panel a}: The frequency of the number of planets detected. A fitted Poisson distribution is shown in red. \textit{Panel b}: The frequency of detected periods, in days. \textit{Panel c}: Frequency of detected planet insolations, as a fraction of the insolation received by Earth ($G_{SC}$). These values were calculated using the Stefan-Boltzmann law and using effective temperatures calculated in Section \ref{sec:overview}. \textit{Panel d}: Frequency of detected planetary radii, in Earth radii. \textit{Panel e}: Frequency of detected transit durations, in hours. \textit{Panel f}: Frequency of detected transit SNRs.}
    \label{fig:summary_quad}
\end{figure*}

We show these normalized distributions in Figure \ref{fig:summary_quad}. In panel \textit{a}, we show a Poisson curve that has been fit to the distribution of planet detections. The Poisson curve has a mean of 1.88$^{+1.44}_{-1.16}$ detections (68\% encompassing width).

The remaining distributions are characterized by their median along with uncertainties based on the 16th and 84th percentiles of each distribution. Panel \textit{b} shows the distribution of detected planet periods, which has a median of 4.00$^{+7.63}_{-3.02}$ days. In panel \textit{c}, we show an estimate of the insolation received by each planet as a fraction of that received by Earth, calculated using the Stefan-Boltzmann law and the effective temperatures derived for each target in Section \ref{sec:overview}. From this, we find that a typical detected planet receives an insolation of 0.36$^{+1.15}_{-0.30}$ times the insolation of Earth. Based on temperature alone, then, planets detected through the survey strategies outlined here are likely to be interesting targets from a habitability perspective, and their atmospheres could potentially be probed for biosignatures with the upcoming \textit{JWST} mission \citep[e.g.,][]{Belu2013}. However, this simple analysis does not account for flaring events around target LTs, which would likely impact prospects for habitability \citep[e.g.,][]{Jackman2019}.




Panel \textit{d} shows the distribution of detected planet radii, with a median of 1.51$^{+0.88}_{-0.44} R_\oplus$. The distribution of transit durations, shown in panel \textit{e}, has a median of 0.81$^{+0.36}_{-0.26}$ hours. This is consistent with the one-hour transit timescale assumed in Section \ref{sec:lc_scatter}. Finally, the SNR of transit detections, shown in panel \textit{f}, has a median value of 19.7$^{+29.9}_{-9.5}$.

\section{Practical Limitations} \label{sec:caveats}
In this section we list a number of limiting factors for performing a search for transiting planets as detailed in Section \ref{sec:simulation}. This is not an exhaustive list, but it represents some of the more prominent limitations to performing a transit survey for planets around LTs in the way we suggested in Section \ref{sec:simulation}.

\subsection{Stability of Dithering Photometry}
\label{sec:stab}
We found in Section \ref{sec:simulation} that an optimal observing strategy with a single telescope requires dithering between multiple objects on each night. As a result, it is critical to maintain photometric precision when switching between targets. For many NIR array detectors, the necessary precision is degraded by imperfect flat-fielding. If a target is placed on a different detector location than it was previously, flat-fielding issues could result in a difference in the measured flux, which could in principle mimic or wipe out the signal of a transiting planet \citep[e.g.,][]{Croll2015}.

\begin{figure*}
    \centering
    \includegraphics[width=\textwidth]{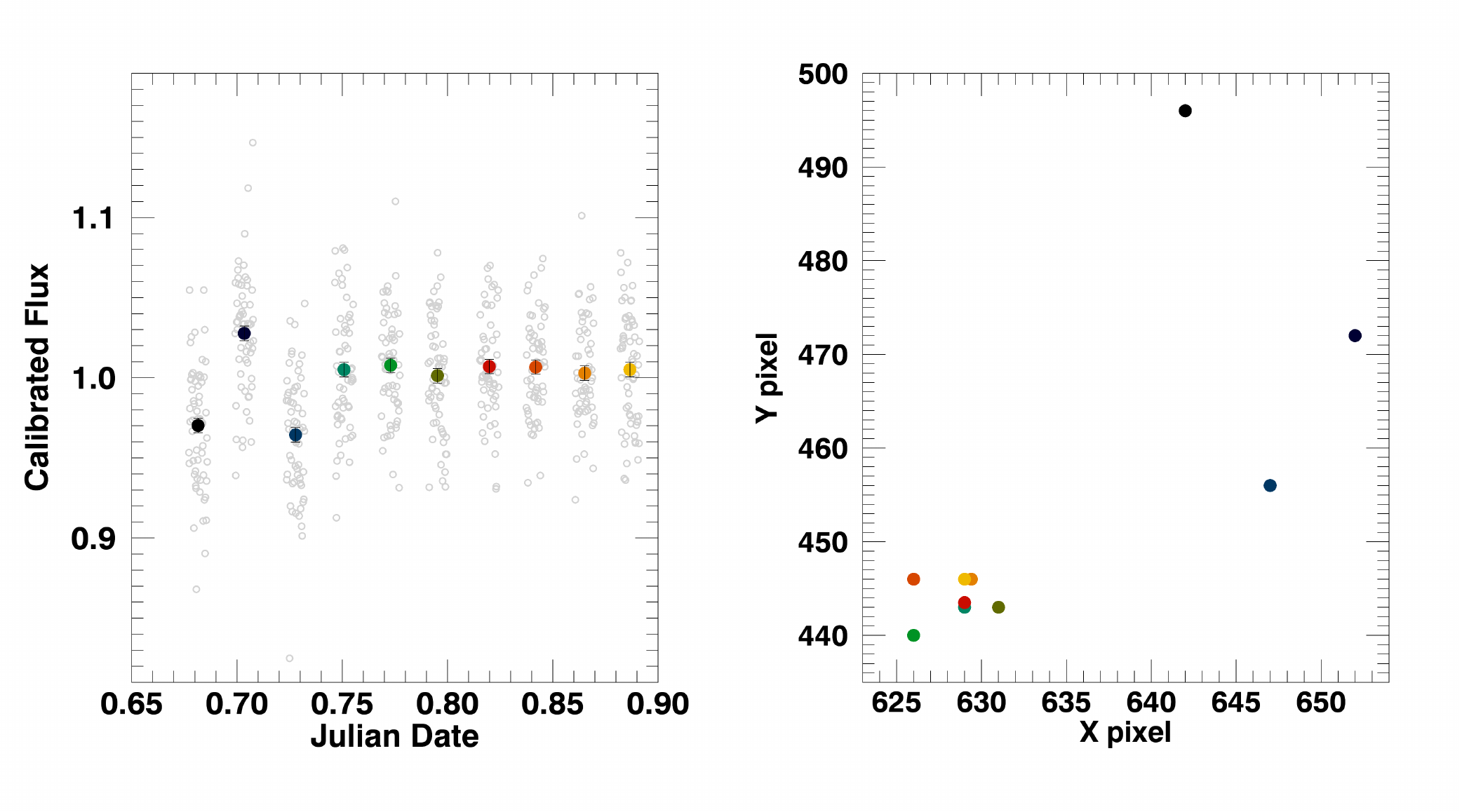}
    \caption{\textit{Left}: One night of \textit{Mimir} photometry of 2MASS 1337, a m$_J$ = 13.8 L0 dwarf, obtained in late May 2018 with the 1.8-m Perkins Telescope. We switched between this object and one other (leading to gaps in the data), in order to test searching multiple targets for planets in one night with a single telescope. \textit{Right}: The centroid position of each block of data, color-coded to match the corresponding block in the left panel. In the first three blocks, the target was placed on different pixels, leading to changes in the normalized flux of over 3$\%$. In the remaining blocks, it was placed in the same area of the detector, which resulted in sub-1$\%$ stability, necessary for finding Earth-sized planets.}
    \label{fig:dithering}
\end{figure*}

In an observing run in May 2018, we tested the effect of ignoring target placement after slews, the result of which is shown in Figure \ref{fig:dithering}. We observed 2MASS 1337, a m$_J$ = 13.8 L0 dwarf, with the \textit{Mimir} instrument on the Perkins Telescope. The target was observed on over five hours on UT 25 May 2016 in \textit{J}-band using a 10-s exposure time. The weather was clear and humidity at 20\%. We switched between 2MASS 1337 and another target every 15 minutes. We found that if we ignored the placement of the target on the detector, the resulting normalized lightcurve showed jumps of over 3$\%$, an effect that would completely swamp the signal of a transiting Earth-sized planet around an LT. When placing the target more carefully, we achieve sub-1$\%$ stability, with the standard deviation of the last seven blocks being 0.2\%. The ability to place a target on a consistent detector location will vary from telescope-to-telescope, but this is a significant source of systematic noise that has to be considered when performing a survey using the dithering strategy we have outlined here. It should be noted that a dithering strategy is employed by the MEarth strategy, which has detected four planets to date.

\subsection{LT Variability}
A non-negligible fraction of LTs show short-term variability over a range of wavelengths, which can in principle complicate the detection of planets, especially in non-continuous lightcurves.
At NIR wavelengths, \citet{Radigan2014b} found that 3.2$^{+2.8}_{-1.8}$\% of targets outside of the L/T transition exhibited large amplitude ($>$2\%) variability, compared to 24$^{+11}_{-9}$\% within it.
Other analyses have inferred a higher rate of low-amplitude variability. \citet{Metchev2015} found evidence that 80$^{+20}_{-27}$\% (95\% confidence level) of L3-L9.5 dwarfs show larger than 0.2\% variability at either 3.6$\mu$m or 4.5$\mu$m, while 36$^{+26}_{-17}$\% of T0-T8 dwarfs show variability larger than 0.4\%.

The timescale of this variability, caused by the rotation of LT targets, is typically in the range of 2 to 12 hours \citep{Artigau2018}. If periodic variability on longer timescales is present, it should be distinguishable from a transit event, which typically last about one hour. The detection of planetary transits in the presence of 1-5\% periodic short-term variability has been demonstrated by \citet{Rizzuto2017}. However, non-periodic variability due to, for example, evolving weather patterns, will complicate transit detection in non-continuous photometry \citep[][]{Gillon2013b}.


\subsection{Single-Transit Detections and Candidate Follow-up}
With the observational strategy outlined in Section \ref{sec:simulation}, most targets are observed for five nights each. Treating the planet population around M dwarfs as representative of that around LTs, most planets have orbital periods longer than five days, so any single planet is unlikely to transit multiple times during a lightcurve. This prevents accurate estimation of the planet's orbital period, and further observation of the target would be required to measure this. Algorithms exist for the estimation of orbital parameters from a single transit \citep[e.g., \textsc{Namaste},][]{Osborn2016}, but whether or not such algorithms perform as well on non-continuous photometry is unclear. 


\subsection{False Positives and Measuring Masses}

Grazing eclipses in binary systems are astrophysical false positives that could mimic the anticipated transit depths of planets around LTs \citep[e.g.,][]{Odonovan2007}. 
The expected number of these false planet detections over the course of an LT transit survey depends on the binary fraction of late-type objects, which is an active area of research. \citet{Bardalez2019} analyzed a sample of 410 spectroscopically-confirmed M7-L5 targets within 25 pc, and measured a spectral binary fraction of 1.6$^{+0.5}_{-0.5}$\%, and a total binary fraction of  7.5$^{+1.6}_{-1.4}$\%. While this sample is not fully complete for L0-L5 dwarfs (83$^{+11}_{-10}$\%), these measurements indicate that LTs occur in binaries less frequently compared to earlier spectral types \citep[e.g.,][ and references therein]{Fontanive2018}. For this reason, astrophysical false positives from eclipsing binaries may occur infrequently over the course of an LT transit survey. Additionally, any new LT eclipsing binaries identified as the result of a transiting planet survey would be interesting in their own right, and would help to further constrain the binary fraction of late-type objects. Regardless, binarity could be ruled out using Doppler observations in the infrared, for example with the Habitable Planet Finder spectrometer on the Hobby-Eberly Telescope \citep{Mahadevan2014} or NIRSpec on Keck II \citep[e.g.][]{Burgasser2016}.

Unfortunately, because LTs are intrinsically faint, measuring planetary masses will not be possible through Doppler measurements with current technology. If the candidate is in a multi-planet system, it may be possible to confirm the planetary nature of orbiting bodies \citep[e.g.][]{Gillon2017,Muirhead2012}; otherwise, candidates can only be considered to be validated, not confirmed.

\section{Conclusions}
We presented considerations for the detection of transiting exoplanets around L and T dwarfs. Despite abundant evidence for the presence of planets around them, these spectral types have not been thoroughly probed for planets. The planetary population around these objects, which span the boundaries between stars, brown dwarfs, and planets themselves, is crucial for further developing our understanding of satellite formation. In addition, transiting Earth-sized planets around these objects will be important targets for follow-up atmospheric characterization with \textit{JWST} \citep{Belu2011}.

By comparing the photometric precision of simulated lightcurves to anticipated transit depths, we showed that the ground-based detection of Earth-sized planets around LTs is facilitated by using a low-dark-current NIR detector. Optical detectors, which typically offer better performance than NIR array detectors, are less efficient for performing a search of transiting planets around LTs, owing to the intrinsic faintness of these objects at  visible wavelengths. 


We also simulated a survey of LT targets in the NIR, which we used to develop an optimal observing strategy with a single telescope. We find that observing multiple targets per night in a dithering strategy, and observing groups for five nights total, delivers the highest number of detected planets, on average.

We performed this simulation for survey durations ranging from one to five years, and telescope sizes ranging from 0.5 to 4.0-m. We recorded the success rate of each combination, and find that a 3-4 year survey with a 2-m class telescope is an efficient design for probing a target sample of LTs. 

Due to large transit depths, planets around Jupiter-sized LTs offer exciting prospects for the characterization of Earth-like atmospheres in the \textit{JWST} era. Such observations would enable comparative planetology studies between our own Earth and others in the universe, placing our own world in a broader context. More optimistically, if they can sustain life, such planets (and others around very low mass stars) will serve as the test bed for the detection of biomarkers with upcoming and future space-based observatories. The detection of these planets is therefore an important first step in the advancement of our understanding of exo-Earths. 

\vspace{5mm}

The authors thank the anonymous referee for comments that improved the quality of this work. The authors also acknowledge Svetlana Jorstad for providing an estimate of sky brightness in \textit{z'}-band. 

\vspace{5mm}
\software{
                \texttt{BATMAN} \citep{Kreidberg2015} }


\newpage
\begin{thebibliography}{}

\bibitem[Alam et al.(2015)]{Alam2015} Alam, S., Albareti, F.~D., Allende Prieto, C., et al.\ 2015, ApJS, 219, 12


\bibitem[Artigau(2018)]{Artigau2018} Artigau, \'E., 2018, Handbook of Exoplanets, eds. Deeg, H. J. \& Belmonte, J. A., DOI: 10.1007/978-3-319-30648-3$\_$94-1


\bibitem[Bailer-Jones, \& Lamm(2003)]{Bailer-Jones2003} Bailer-Jones, C.~A.~L., \& Lamm, M.\ 2003, MNRAS, 339, 477

\bibitem[Ballard \& Johnson(2016)]{Ballard2016} Ballard, S. \& Johnson, J. A. 2016, ApJ, 816, 66

\bibitem[Baraffe et al.(2015)]{Baraffe2015} Baraffe, I., Homeier, D., Allard, F., et al. 2015, A\&A, 577, A42

\bibitem[Bardalez Gagliuffi et al.(2019)]{Bardalez2019}Bardalez Gagliuffi, D.~C., Burgasser, A.~J., Schmidt, S.~J., et al. 2019, arXiv e-prints, arXiv:1906.04166

\bibitem[Barnes \& Heller(2013)]{Barnes2013} Barnes, R. \& Heller, R. 2013, AsBio, 13, 279

\bibitem[Belu et al.(2011)]{Belu2011} Belu, A. R., Selsis, F.,  Morales, J.-C. et al., 2011, A\&A, 525, A83

\bibitem[Belu et al.(2013)]{Belu2013} Belu, A.~R., Selsis, F., Raymond, S.~N., et al. 2013, ApJ, 768, 125


\bibitem[Berta et al.(2012)]{Berta2012} Berta, Z.~K., Irwin, J., Charbonneau, D., et al.\ 2012, AJ, 144, 145


\bibitem[Berta-Thompson et al.(2015)]{BertaThompson2015} Berta-Thompson, Z. K., Irwin, J., Charbonneau, D. et al., 2015, Nature, 527, 204


\bibitem[Blake et al.(2008)]{Blake2008} Blake, C.~H., Bloom, J.~S., Latham, D.~W., et al.\ 2008, PASP, 120, 860

\bibitem[Bolmont et al.(2017)]{Bolmont2017} Bolmont, E., Selsis, F., Owen, J. E. et al., 2017, MNRAS, 464, 3728

\bibitem[Borucki et al.(2010)]{Borucki2010} Borucki, W. J., Koch,  D., Basri, G. et al., 2010, Science, 327, 5968

\bibitem[Burdanov et al.(2017)]{Burdanov2017} Burdanov, A., Delrez, L, Gillon, M. et al., 2017, SPECULOOS Exoplanet Search and Its Prototype on TRAPPIST, p.130, doi:10.1007/978-3-319-30648-3$\_$130-1

\bibitem[Burgasser(2001)]{Burgasser2001} Burgasser, A.~J.\ 2001, Ph.D. Thesis

\bibitem[Burgasser et al.(2006)]{Burgasser2006} Burgasser, A. J., Geballe, T. R., Leggett, S. K. et al., 2006, ApJ, 637, 1067

\bibitem[Burgasser et al.(2016)]{Burgasser2016} Burgasser, A.~J., Blake, C.~H., Gelino, C.~R., et al.\ 2016, ApJ, 827, 25


\bibitem[Burrows et al.(1997)]{Burrows1997} Burrows, A., Marley, M. Hubbard, W. B. et al., 1997, ApJ, 491, 856

\bibitem[Burrows et al.(2011)]{Burrows2011} Burrows, A., Heng, K., \& Nampaisarn, T. 2011, ApJ, 736, 47

\bibitem[Charbonneau et al.(2000)]{Charbonneau2000} Charbonneau, D., Brown, T. M., Latham, D. W., \& Mayor, M. 2000, ApJL, 529, L45

\bibitem[Charbonneau et al.(2002)]{Charbonneau2002} Charbonneau, D., Brown T.  M., Noyes R. W., \& Gilliland R. L. 2002, ApJ, 568, 377

\bibitem[Charbonneau et al.(2009)]{Charbonneau2009} Charbonneau, D., Berta, Z. K., Burke, C. J. et al., 2009, Nature, 462, 891

\bibitem[Chauvin et al.(2004)]{Chauvin2004} Chauvin, G., Lagrange, A.-M., Dumas, C., et al.\ 2004, A\&A, 425, L29

\bibitem[Chauvin et al.(2005)]{Chauvin2005} Chauvin, G., Lagrange, A.-M., Dumas, C., et al.\ 2005, A\&A, 438, L25




\bibitem[Claret, \& Bloemen(2011)]{Claret2011} Claret, A., \& Bloemen, S.\ 2011, A\&A, 529, A75


\bibitem[Clemens et al.(2007)]{Clemens2007} Clemens, D. P., Sarcia, D., Grabau, A. et al., 2007, PASP, 119, 862

\bibitem[Cohen et al.(2003)]{Cohen2003} Cohen, M., Wheaton, W.~A., \& Megeath, S.~T.\ 2003, AJ, 126, 1090

\bibitem[Croll et al.(2015)]{Croll2015} Croll, B., Albert, L., Jayawardhana, R. et al., 2015, ApJ, 802, 28

\bibitem[Cutri et al.(2003)]{Cutri2003} Cutri, R.~M., Skrutskie, M.~F., van Dyk, S., et al.\ 2003, VizieR Online Data Catalog, II/246


\bibitem[Delrez et al.(2018)]{Delrez2018} Delrez, L., Gillon, M., Queloz, D. et al., 2018, Proc. SPIE 10700, Ground-based and Airborne Telescopes VII, 107001I 

\bibitem[Dittmann et al.(2017)]{Dittmann2017} Dittmann, J. A., Irwin, J. M., Charbonneau, D. et al., 2017, Nature, 544, 333

\bibitem[Downes et al.(2015)]{Downes2015} Downes, J. J., Rom\'{a}n-Z\'{u}\~{n}iga, C., Ballesteros-Paredes, J. et al., 2015, MNRAS, 450, 3490


\bibitem[Dressing \& Charbonneau(2015)]{Dressing2015} Dressing, C. D. \& Charbonneau, D. 2015, ApJ, 807, 45


\bibitem[Dupuy et al.(2014)]{Dupuy2014} Dupuy, T.~J., Liu, M.~C., \& Ireland, M.~J.\ 2014, ApJ, 790, 133






\bibitem[Faherty et al.(2016)]{Faherty2016} Faherty, J.~K., Riedel, A.~R., Cruz, K.~L., et al. 2016, ApJS, 225, 10


\bibitem[Fontanive et al.(2018)]{Fontanive2018} Fontanive, C., Biller, B., Bonavita, M., et al.\ 2018, MNRAS, 479, 2702


\bibitem[Fukugita et al.(1996)]{Fukugita1996} Fukugita, M., Ichikawa, T., Gunn, J.~E., et al.\ 1996, AJ, 111, 1748

\bibitem[Gagne(2014)]{Gagne2014} Gagne, J. 2014, \textit{List of All UltraCool Dwarfs}, https://jgagneastro.wordpress.com/list-of-ultracool-dwarfs/

\bibitem[Gillon(2013a)]{Gillon2013} Gillon, M., Jehin, E., Fumel, A. et al., 2013, EPJ Web Conf. 47, 03001 (2013)

\bibitem[Gillon et al.(2013b)]{Gillon2013b} Gillon, M., Triaud, A.~H.~M.~J., Jehin, E., et al.\ 2013, A\&A, 555, L5


\bibitem[Gillon et al.(2016)]{Gillon2016} Gillon, M., Jehin, E., Lederer, S.~M., et al.\ 2016, Nature, 533, 221.

\bibitem[Gillon et al.(2017)]{Gillon2017} Gillon, M., Triaud, A. H., Demory, B.-O. et al., 2017, Nature, 542, 456

\bibitem[Han et al.(2016)]{Han2016} Han, C, Bennett, D. P., Udalski, A., \& Jung, Y. K. 2016, ApJ, 825, 1

\bibitem[Hardegree-Ullman et al.(2019)]{Hardegree2019} Hardegree-Ullman, K.~K., Cushing, M.~C., Muirhead, P.~S., et al.\ 2019, arXiv e-prints, arXiv:1905.05900

\bibitem[Harding et al.(2016)]{Harding2016} Harding, L. K., Hallinan, G., Milburn, J. et al., 2016, MNRAS, 457, 3

\bibitem[He et al.(2017)]{He2017} He, M. Y., Triaud, A. H. M. J., \& Gillon, M. 2017, MNRAS, 464, 2687


\bibitem[Ida \& Lin(2004)]{Ida2004} Ida, S. \& Lin, D. N. C. 2004, ApJ, 603, 388


\bibitem[Irwin et al.(2011)]{Irwin2011} Irwin, J., Berta, Z.~K., Burke, C.~J., et al.\ 2011, ApJ, 727, 56

\bibitem[Jackman et al.(2019)]{Jackman2019} Jackman, J.~A.~G., Wheatley, P.~J., Bayliss, D., et al. 2019, MNRAS, 485, L136

\bibitem[Jenkins et al.(2002)]{Jenkins2002} Jenkins, J.~M., Caldwell, D.~A., \& Borucki, W.~J.\ 2002, ApJ, 564, 495


\bibitem[Jung et al.(2018)]{Jung2018} Jung, Y. K., Udalski, A., Gould, A. et al., 2018, AJ, 155, 5

\bibitem[Kirkpatrick et al.(1997)]{Kirkpatrick1997} Kirkpatrick, J. D., Henry, T. J., \& Irwin, M. J. 1997, AJ, 113, 4
 
\bibitem[Kirkpatrick et al.(1999)]{Kirkpatrick1999} Kirkpatrick, J. D., Reid, I. N., Liebert, J. et al., 1999, ApJ, 519, 802

\bibitem[Kreidberg(2015)]{Kreidberg2015} Kreidberg, L. 2015, PASP, 127, 1161


\bibitem[Luger et al.(2017)]{Luger2017} Luger, R., Lustig-Yaeger, J., \& Agol, E.\ 2017, ApJ, 851, 94


\bibitem[Luhman \& Mamajek(2012)]{Luhman2012} Luhman, K. L. \& Mamajek, E. E. 2012, ApJ, 758, 1


\bibitem[Mahadevan et al.(2014)]{Mahadevan2014} Mahadevan, S., Ramsey, L. W., Terrien, R. et al., 2014, Proc. SPIE, 9147, 91471G

\bibitem[Mamajek(2005)]{Mamajek2005} Mamajek, E.~E.\ 2005, ApJ, 634, 1385

\bibitem[Ment et al.(2019)]{Ment2019} Ment, K., Dittmann, J.~A., Astudillo-Defru, N., et al.\ 2019, AJ, 157, 32

\bibitem[Metchev et al.(2015)]{Metchev2015} Metchev, S. A., Heinze, A., Apai, D. et al., 2015, ApJ, 799, 154 

\bibitem[Morley et al.(2017)]{Morley2017} Morley, C. V., Kreidberg, L., Rustamkulov, Z. et al., 2017, ApJ, 850, 121

\bibitem[Muirhead et al.(2012)]{Muirhead2012} Muirhead, P. S., Johnson, J. A., Apps, K. et al., 2012, ApJ, 747, 144

\bibitem[Mulders et al.(2015)]{Mulders2015} Mulders, G.~D., Pascucci, I., \& Apai, D.\ 2015, ApJ, 814, 130

\bibitem[Nutzman \& Charbonneau(2008)]{Nutzman2008} Nutzman, P. \& Charbonneau, D. 2008, PASP, 120, 317

\bibitem[O'Donovan et al.(2007)]{Odonovan2007} O'Donovan, F.~T., Charbonneau, D., Alonso, R., et al.\ 2007, ApJ, 662, 658


\bibitem[Osborn et al.(2016)]{Osborn2016} Osborn, H. P., Armstrong, D. J., Brown, D. J. A. et al., 2016, MNRAS, 457, 2273

\bibitem[Payne \& Lodato(2007)]{Payne2007} Payne, M. J. \& Lodato, G. 2007, MNRAS, 381, 1597

\bibitem[Pollacco et al.(2006)]{Pollacco2006} Pollacco, D. L., Skillen, I., Collier Cameron, A. et al., 2006, PASP, 118, 848



\bibitem[Radigan et al.(2014)]{Radigan2014a} Radigan, J., Jayawardhana, R., Lafreniere, D. et al., 2014, ApJ, 750, 105

\bibitem[Radigan (2014)]{Radigan2014b} Radigan, J. 2014, ApJ, 797, 2

\bibitem[Ricker et al.(2015)]{Ricker2015} Ricker, G. R., Winn, J. N., Vanderspek, R. et al., 2015, Journal of Astronomical Telescopes, Instruments, and Systems, 1, 014003

\bibitem[Rizzuto et al.(2017)]{Rizzuto2017} Rizzuto, A. C., Mann, A. W., Vanderburg, A. et al., 2017, AJ, 154, 6

\bibitem[Ross \& Schubert(1987)]{Ross1987} Ross, M. N. \& Schubert, G. 1987, Nature, 325, 6100

\bibitem[Ross \& Schubert(1989)]{Ross1989} Ross, M. N., \& Schubert, G. 1989, Icarus, 78, 1



\bibitem[Seager et al.(2007)]{Seager2007} Seager, S., Kuchner, M., Hier-Majumder, C.~A., et al.\ 2007, ApJ, 669, 1279




\bibitem[Sullivan et al.(2014)]{Sullivan2014} Sullivan, P. W., Croll, B. \& Simcoe, R. A. 2014, Proc. SPIE, 9154, 91541F

\bibitem[Sullivan et al.(2015)]{Sullivan2015} Sullivan, P.~W., Winn, J.~N., Berta-Thompson, Z.~K., et al.\ 2015, ApJ, 809, 77



\bibitem[Udalksi et al.(2015)]{Udalski2015} Udalski, A., Jung, Y. K., Han, C. et al., 2015, ApJ, 812, 1

\bibitem[Van Grootel et al.(2018)]{VanGrootel2018} Van Grootel, V., Fernandes, C. S., Gillon, M. et al., 2018, ApJ, 853, 30


\bibitem[Yoder(1979)]{Yoder1979} Yoder, C. F. 1979, Nature, 279, 5716


\end{thebibliography}
\end{document}